\documentstyle[aps,preprint]{revtex}
\newcommand\beq{\begin{equation}}
\newcommand\eeq{\end{equation}}
\newcommand\bea{\begin{eqnarray}}
\newcommand\eea{\end{eqnarray}}
\newcommand\pa{\partial}
\newcommand\un{\underline}

\newcommand\pr{\prime}

\begin{document}
\draft
\preprint{\vbox{\hbox{HD-THEP-01-46, SOGANG-HEP 289/01}}}
\title{Symplectic embedding and Hamilton-Jacobi analysis of Proca model}
\author{Soon-Tae Hong$^{1}$, Yong-Wan Kim$^{1}$, Young-Jai Park$^{1}$ and 
K. D. Rothe$^{1,2}$}
\address{$^{1}$Department of Physics 
and Basic Science Research Institute,\\
Sogang University, C.P.O. Box 1142, Seoul 100-611, Korea\\
$^{2}$Institut f\"ur Theoretische Physik,\\
Universit\"at Heidelberg, Philosophenweg 16, D-69120 Heidelberg, Germany\\}
\date{\today}
\maketitle

\begin{abstract}
 
Following the symplectic approach we show how to embed 
the Abelian Proca model into a first-class system 
by extending the configuration space to include an additional pair of 
scalar fields, and compare it with the improved Dirac scheme.  
We obtain in this way the desired Wess-Zumino and gauge fixing terms of 
BRST invariant Lagrangian.  Furthermore, the integrability properties of 
the second-class system described by the Abelian Proca model are investigated 
using the Hamilton-Jacobi formalism, where we construct the closed Lie algebra 
by introducing operators associated with the generalized Poisson 
brackets.\\ \\ 
\noindent
PACS number(s):11.10.-z, 11.10.Ef, 11.15.Tk, 11.30.-j\\
\noindent
Keywords: Proca model, symplectic embedding, Hamilton-Jacobi scheme 
\end{abstract}

\newpage

\section{Introduction}

The standard Dirac quantization method (DQM)~\cite{dirac64} has been
widely used in order to quantize Hamiltonian systems
involving first- and second-class constraints.
However, the resulting Dirac brackets
may be field-dependent and nonlocal, and thus pose serious ordering problems
for the quantization of the theory.  On the other hand, the
Becci-Rouet-Stora-Tyutin (BRST)~\cite{becci76,kugo79}
quantization of constrained systems along the lines originally
established by Batalin, Fradkin, and Vilkovisky
~\cite{fradkin75,henneaux85}, and then reformulated
in a more tractable and elegant version by
Batalin, Fradkin, and Tytin~\cite{batalin87},
does not suffer from these difficulties,
as it relies on a simple Poisson bracket structure. As a result,
the embedding of second-class systems into first-class ones
(gauge theories) has received much attention in the past years and the
DQM improved in this way, has been applied to a number of
models~\cite{idqm,nonPro,gafn,fujiwara90,kim92,banerjee94,phyrep}
in order to obtain the corresponding Wess-Zumino (WZ)
actions~\cite{faddeev86,wess71}.  In fact, the earlier work on this
subject is based on the traditional Dirac's pioneering
work~\cite{dirac64}, which has been criticized for introducing
``superfluous'' primary constraints, and has been avoided in more recent
treatments, based on the symplectic structure of phase space.
That this approach is of particular advantage in the case of first-order
Lagrangians such as Chern-Simons theories has been emphasized by
Faddeev and Jackiw~\cite{jackiw85}. This symplectic scheme has been applied
to a number of models~\cite{wozneto,kimjkps1} and has recently been used to
implement the improved DQM embedding program in the context of the symplectic
formalism~\cite{kimjkps2,neto0109}.

Based on the Carath${\rm\acute e}$odory equivalent Lagrangians
method~\cite{car}, an alternative Hamilton-Jacobi (HJ) scheme for constrained
systems was also proposed~\cite{guler89} and exploited to
quantize singular systems~\cite{pimentel98,baleanu01,hong01ph}.
One of the most interesting applications of the HJ
scheme is system with second-class
constraints~\cite{dirac64,gomis}, since the set of differential equations
derived from the corresponding HJ equation is not integrable~\cite{gomis},
being incomplete.  They become complete with the addition of suitable 
``integrability conditions," which turn out to be Dirac ``consistency
conditions" requiring time independence of the constraints~\cite{hong01ph}.

In this paper, we wish to illustrate the above quantization schemes in the
case of the Abelian Proca model.  The material is organized as follows.
In section 2, we briefly recapitulate the Proca model in the framework of the
standard and the improved DQMs. In section 3, after briefly
reviewing the gauge non-invariant symplectic formalism for this
model~\cite{jackiw85}, we then show how the improved DQM program for embedding
this second-class system into a first-class one is realized in the
framework of the symplectic formalism, and obtain in this way the
corresponding Wess-Zumino and gauge fixing terms of the BRST invariant 
Lagrangian.  In section 4, we finally apply the HJ quantization scheme to 
the Proca model and comment on the integrability conditions and the closed 
Lie algebra obtained by introducing operators associated with the 
generalized Poisson brackets.  Our conclusion is given in section 5.

\section{Dirac quantization method}
\setcounter{equation}{0}
\renewcommand{\theequation}{\arabic{section}.\arabic{equation}}
\begin{center}
{\bf Standard Dirac quantization method}
\end{center}

In this section, we briefly recapitulate the massive Proca model described 
by the Lagrangian  
\begin{equation}
\label{action}
{\cal L}_{0}= -\frac{1}{4} F_{\mu\nu} F^{\mu\nu}
             + \frac{1}{2} m^2 A_\mu A^{\mu},
\end{equation}
where $F_{\mu\nu} = \partial_\mu A_\nu - \partial_\nu A_\mu$ and
$g_{\mu\nu} = {\rm diag}(+,-,-,-)$.  
The canonical momenta conjugate to the fields $A^{\mu}$ are given by
$\pi_{0} = 0$ and $\pi_{i}= F_{i0}$
with the Poisson algebra $\{ A^{\mu}(x), \pi_{\nu}(y) \}
=\delta ^{\mu}_{\nu} \delta(x-y)$. 
The canonical Hamiltonian then reads
\begin{equation}
\label{canH}
{\cal H}_0=\frac{1}{2}\pi_i^2+\frac{1}{4}F_{ij}F^{ij}
+\frac{1}{2}m^2(A^0)^2 +\frac{1}{2}m^2(A^i)^2-A^0(\partial^i\pi_i+m^2A^0).
\end{equation}
Since we have one primary constraint 
\begin{equation}
\Omega_{1}=\pi_0 \approx 0,
\label{constr0}
\end{equation}
the total Hamiltonian is given by
\begin{equation}
{\cal H}_T={\cal H}_0+u \Omega_{1}
\label{htotal}
\end{equation}
with $u$ a Lagrange multiplier. 
The requirement $\dot{\Omega}_1\approx 0$ leads to the secondary, 
Gauss' law constraint
\begin{equation}
\Omega_2 =\partial^i \pi_i + m^2 A^0 \approx 0.
\label{constr1}
\end{equation}
Note that the time evolution of this constraint with ${\cal H}_T$ generates 
no further constraint, but only fixes the multiplier $u$ to be 
$u=-\partial_i A^i$, so that ${\cal H}_T$ no longer involves arbitrary parameters:
\begin{equation}
{\cal H}_T={\cal H}_0-\Omega_{1}\partial_i A^{i}.
\label{htotal1}
\end{equation}
As a result, the full set of constraints of this
model is $\Omega_{i}$ $(i,j = 1, 2)$. They satisfy the second-class 
constraint algebra
\begin{equation}
\Delta_{ij}(x,y)=\{ \Omega_i(x), \Omega_j(y) \}
               = - m^2 \epsilon_{ij} \delta(x-y)
\end{equation}
with $\epsilon_{12}=-\epsilon_{21}=1$.
The consistent quantization of the Proca model is then  
obtained in terms of the Dirac brackets~\cite{kkky} 
\begin{equation}
\begin{array}{ll}
\{A^0 (x), A^j (y) \}_{D}=\frac{1}{m^{2}} \partial_{x}^{j}
\delta(x-y),  
&\{A^0 (x), A^0 (y) \}_{D}=0,\\
\{A^j (x), A^k (y) \}_{D}=0,
&\{\pi_\mu (x), \pi_\nu (y) \}_{D}=0,\\
\{A^i (x), \pi_j (y) \}_{D}=\delta^{i}_{j} \delta (x-y), 
&\{A^0 (x), \pi_{\nu} (y) \}_{D}=0,\\
\{A^i (x), \pi_0 (y) \}_{D}=0.
&  \\
\end{array}
\label{diracs}
\end{equation}

For later comparison we also list the equations of motion
following from the time evolution of the fields $A^{\mu}$ and $\pi_{\mu}$ with 
${\cal H}_T$:
\begin{equation}
\begin{array}{ll}
\dot{A_{0}}= -\partial_i A^i ,
&\dot{A}^{i}=\pi_i+\partial^iA^0,\\
\dot{\pi}_{0}=\partial^i\pi_i+m^2A^0, 
&\dot{\pi}_{i}=\partial_j F^{ij}-m^2A^{i},\\
\end{array}
\label{pqdots}
\end{equation}
which, together with the constraints $\Omega_{i}$ , 
reproduce the well-known equations 
\begin{equation}
(\partial_\nu\partial^\nu+m^2)A^\mu=0.
\label{da}
\end{equation}

\begin{center}
{\bf Improved Dirac Quantization Method}
\end{center}

For late comparison we now briefly review the improved 
DQM~\cite{idqm,nonPro,gafn,fujiwara90}, which implements the conversion of the 
second-class constraints of a system~\cite{batalin87} to the first-class 
constraints, for the case of the Abelian Proca model~\cite{banerjee95ap,kkky}.  
To this end we extend phase space by introducing a pair of auxiliary fields 
$(\theta,\pi_{\theta})$ satisfying the canonical Poisson brackets 
\begin{equation}
\{\theta (x),\pi_{\theta}(y) \}=\delta(x-y).
\label{auxil}
\end{equation}

Following the improved DQM, 
we obtain for the Abelian conversion of the second-class constraints,
Eqs. (\ref{constr0}) and (\ref{constr1}), to the first-class constraints
\begin{equation}
\tilde{\Omega}_{1}=\pi_{0}+m^{2}\theta,~~~
\tilde{\Omega}_{2}=\partial^{i}\pi_{i}+m^{2}A^{0}+\pi_{\theta}
\label{consttil}
\end{equation}
satisfying the rank-zero algebra
\begin{eqnarray}
\{\tilde{\Omega}_{i}, \tilde{\Omega}_{j} \}=0.
\label{omegastil}
\end{eqnarray}

Similarly, we obtain for the first-class physical fields in the extended phase 
space
\begin{equation}
\tilde{A}^{\mu}=(A^{0}+\frac{1}{m^{2}}\pi_{\theta}, A^{i}
+\partial^{i}\theta),~~~
\tilde{\pi}_{\mu}=(\pi_{0}+m^{2}\theta, \pi_{i}).
\label{apitilde} 
\end{equation}
Since an arbitrary functional of the first-class physical fields is also 
first-class~\cite{idqm}, we can directly obtain the desired 
first-class Hamiltonian $\tilde{\cal H}$ corresponding to the Hamiltonian 
${\cal H}_T$ in Eq. (\ref{htotal1}) via the substitution 
$A^\mu\rightarrow\tilde{A}^\mu$, $\pi_\mu\rightarrow\tilde{\pi}_\mu$:
\begin{equation}
\tilde{\cal H}={\cal H}_T+
  \frac{1}{2}m^{2}(\partial_{i}\theta)^{2}
 +\frac{1}{2m^{2}}\pi_{\theta}^{2}
 +\partial_{i}^{2}\theta\tilde{\Omega}_{1}
 -\frac{1}{m^{2}}\pi_{\theta}\tilde{\Omega}_{2}. 
\label{htotalt}
\end{equation}

On the other hand, one easily recognizes that the Poisson brackets 
between the first-class fields in the extended phase space  
are formally identical with the Dirac brackets of the corresponding 
second-class fields~\cite{gr}.  Note that the symplectic 
formalism~\cite{jackiw85,wozneto} also gives the same result.  
(See next section for more details.)

Next, one can consider the partition function of the model
in order to present the Lagrangian corresponding to $\tilde{\cal H}$ in
the canonical Hamiltonian formalism as follows
\begin{equation}
Z= \int{\cal D}A^\mu {\cal D}\pi_\mu {\cal D}\theta
       {\cal D}\pi_{\theta}\prod_{i,j =1}^{2} 
       \delta(\tilde{\Omega}_i)\delta(\Gamma_j)
       \mbox{det}\mid\{\tilde{\Omega}_i, \Gamma_j \} \mid
       e^{iS},
\end{equation}
where
\begin{equation}
S=\int {\rm d}^4 x\left(\pi_\mu {\dot A}^\mu 
+ \pi_\theta \dot{\theta}-\tilde{\cal H}\right),
\end{equation}
with the Hamiltonian $\tilde{H}$ in Eq. (\ref{htotalt}).  
Then, after performing tedious integrations over all momenta,
one obtains the Lagrangian
\begin{equation}
{\cal L}=-\frac{1}{4}  F_{\mu\nu} F^{\mu\nu}
+\frac{1}{2}m^{2}(A_{\mu}+\partial_{\mu}\theta)
(A^{\mu}+\partial^{\mu}\theta).
\label{stuck}
\end{equation}
Up to a total divergence term this is just the manifestly gauge invariant 
 St\"uckelberg Lagrangian, with $\theta$ the St\"uckelberg scalar, 
which is invariant under the gauge transformations as
$\delta A^{\mu} = \partial^{\mu} \Lambda$ and $\delta \theta = - \Lambda$.

\section{Symplectic Embedding Formalism}
\setcounter{equation}{0}
\renewcommand{\theequation}{\arabic{section}.\arabic{equation}}
\begin{center}
{\bf Gauge noninvariant symplectic scheme}
\end{center}

In order to set the stage for the symplectic embedding of the Proca model 
into a gauge theory, we briefly review~\cite{jackiw85,kkky} 
the gauge noninvariant symplectic formalism for this model. 
Following ref.~\cite{jackiw85}, we rewrite 
the second-order Lagrangian (\ref{action}) as the first-order Lagrangian
\begin{equation}
{\cal L}_{0} = \pi_{0}\dot{A}^{0}+\pi_i \dot{A}^i - {\cal H}_{0},
\label{lagfirst}
\end{equation}
where the Lagrangian ${\cal L}_0$ is to be regarded as a function of
the configuration--space variables $A^i$ and $\pi_i$, and
${\cal H}_0$ is the canonical Hamiltonian density in Eq. (\ref{canH}). 
Since ${\cal H}_0$
depends on $\pi_{i}$ and $A^{i}$, but not on their time derivatives,  
it can be regarded as the (level zero) symplectic potential ${\cal H}^{(0)}$.  

In order to find the symplectic brackets
we introduce the sets of the symplectic variables 
$\xi^{(0)\alpha}=(\vec{A}, \vec{\pi}, A^0)$ and
their conjugate momenta $a^{(0)}_{\alpha}=(\vec{0},\vec{\pi}, 0)$,
which are directly read off from the canonical sector of
the first-order Lagrangian (\ref{lagfirst}),
written in the form 
\begin{equation}
{\cal L} = a^{(0)}_{\alpha} \xi^{(0)\alpha} - {\cal H}^{(0)}.
\label{genL}
\end{equation}  
The dynamics of the model is then governed by the symplectic two-form matrix: 
\begin{equation}
f^{(0)}_{\alpha\beta} (x,y) = \frac{\partial a^{(0)}_\beta(y)}
             {\partial\xi^{(0)\alpha}(x)} -
         \frac{\partial a^{(0)}_\alpha(x)}{\partial \xi^{(0)\beta}(y)},
\end{equation}
via the equations of motion
\begin{equation}
\int{\rm d}^3y f^{(0)}_{\alpha\beta}(x,y)\dot{\xi}^{\beta}(y)=
              \frac{\delta}{\delta \xi^{(0)\alpha}(x)}\int {\rm d}^3 y {\cal H}_{0}(y).
\end{equation}
In the Proca model the symplectic two-form matrix is given by
\begin{equation}
f^{(0)}_{\alpha\beta}(x,y) =
\left( 
\begin{array}{ccc}
O &-I &\vec{0}  \\
I &O &\vec{0} \\
\vec{0}^{T} &\vec{0}^{T} &0  \\
\end{array}
\right)
\delta (x-y),
\end{equation}
where $O (I)$, $\vec{b}$ and $\vec{b}^T$ stand for a $3\times 3$ 
null (identity) matrix, a column vector and its transpose, respectively, 
showing the matrix $f^{(0)}_{\alpha\beta}(x,y)$ is singular.  Here the 
symplectic two-form matrix has a zero mode
\footnote{We label the zero modes as follows: 
$\nu^{(l)}_{\alpha,y}(\sigma,x),~(\sigma=1,...,N)$,
where ``$l$'' refers to the ``level'',
$\alpha$, $y$ stand for the component, while $\sigma$, $x$ label 
the N-fold infinity of zero modes in ${\cal R}^3$.
For simplicity we refer to the zero modes only according to their discrete
labelling $\sigma$.}: 
$\nu^{(0)T}_{\alpha,y}(1,x)=(\vec{0}, \vec{0}, 1)\delta (x-y)$, 
which generates the constraint
$\Omega_{2}$ in the context of the symplectic formalism~\cite{jackiw85} 
as follows
\begin{equation}
\int {\rm d}^3y~\delta (x-y)\frac{\delta}{\delta A^0(y)}
               \int {\rm d}^3 z {\cal H}_{0}(z) 
=-\Omega_{2}(x)=0,
\end{equation}
where $\Omega_{2}$ is given by Eq. (\ref{constr1}).  Following the symplectic 
algorithm, we add the constraint $\Omega_{2}$ to the 
canonical sector of the Lagrangian (\ref{lagfirst}), by enlarging 
the symplectic phase space with the addition of 
a Lagrange multiplier $\rho$.  The once iterated first-label Lagrangian 
is then given as
\begin{equation}
{\cal L}^{(1)} = \pi_{0}\dot{A}^{0}+\pi_i \dot{A}^i 
+\Omega_{2}\dot{\rho}-{\cal H}^{(1)},
\label{lagsecond}
\end{equation}
where the first-iterated Hamiltonian 
${\cal H}^{(1)}={\cal H}^{(0)}|_{\Omega_{2}=0}$ is  
given by 
\begin{equation}
{\cal H}^{(1)}=\frac{1}{2}\pi_{i}^2+ \frac{1}{4}F_{ij}F^{ij}
               +\frac{1}{2} m^2 (A^0)^2+ \frac{1}{2} m^2 (A^i)^{2}.
\label{hamilsecond}
\end{equation}

The situation at this stage is exactly the same as before except for the 
replacement, ${\cal L}_{0}\rightarrow{\cal L}^{(1)}$ and 
${\cal H}_{0}\rightarrow{\cal H}^{(1)}$.  In other words, we now have 
for the symplectic variables and their conjugate momenta
\begin{equation}
\xi^{(1)\alpha}=(\vec{A}, \vec{\pi}, A^0, \rho),~~~
a^{(1)}_\alpha =(\vec{\pi}, \vec{0},0, \Omega_{2}).
\end{equation}
The first iterated symplectic nonsingular two-form matrix is now given by
\begin{equation}
f^{(1)}_{\alpha\beta}(x,y) =
\left( \
\begin{array}{cccc}
         O &-I   &\vec{0}  &\vec{0} \\
         I &O  &\vec{0}    &-\vec{\nabla}_x \\
         \vec{0}^T &\vec{0}^T   &0     &m^2 \\
         \vec{0}^T &  \vec{\nabla}^{T}_{x} & -m^2 & 0
\end{array}
\right)
\delta (x-y).
\end{equation}
Its inverse matrix is easily 
obtained
\begin{equation}
(f^{(1)}_{\alpha\beta})^{-1} (x,y) =
\left( 
\begin{array}{cccc}
O  &I &\frac{1}{m^2}\vec{\nabla}_x &  \vec{0} \\
-I &O   & \vec{0}    &  \vec{0} \\
-\frac{1}{m^2}\vec{\nabla}^T_x & \vec{0}^{T}  &0   &  -\frac{1}{m^2} \\
\vec{0}^{T} &\vec{0}^{T} & \frac{1}{m^2} & 0
\end{array}
\right)
\delta (x-y).
\end{equation}
Now, this inverse symplectic
two-form matrix gives the symplectic brackets of the Proca model
\begin{equation}
\{\xi^{(1)\alpha}(x), \xi^{(1)\beta}(y) \}_{symp} 
= (f^{(1)})^{-1}_{\alpha\beta} (x,y),
\end{equation}
which are recognized to be identical with  the Dirac brackets in Eq. 
(\ref{diracs}).

\begin{center}
{\bf Gauge invariant symplectic embedding}
\end{center}

Embedding a second-class system into a first-class one is, 
on Lagrangian level, equivalent to finding the Wess-Zumino(WZ) action
for the Lagrangian in question. This is what we propose to do next 
in the context of the symplectic formalism, taking the Proca model 
as an illustration. The starting point is provided by the Lagrangian
\begin{equation}
{\cal L}=-\frac{1}{4}F_{\mu\nu}F^{\mu\nu}+\frac{1}{2}m^2 A_\mu
         A^\mu + {\cal L}_{WZ}.
\end{equation}
The symplectic procedure is greatly simplified, if we make the following 
``educated guess'' for the WZ Lagrangian, respecting Lorentz symmetry,
\begin{equation}
{\cal L}_{WZ}= \frac{1}{2}c_1 \partial_\mu\theta\partial^\mu\theta
         +c_2 A^{\mu}\partial_\mu \theta + c_3 f,
\label{lagwz}
\end{equation}
with $f$ an arbitrary polynomial of $\theta$.  
As an Ansatz we shall take $c_{i}$ ($i=1,2,3$) to be constants to be fixed 
by the symplectic embedding procedure.  
After partial integration of the second term in Eq. (\ref{lagwz}) in order to 
coincide with the constraint $\tilde{\Omega}_{1}$, in terms of the canonical momenta conjugate to $A^0$, $A^i$ and $\theta$ 
\begin{equation}
\pi_0=-c_{2}\theta,~~~
\pi_i=F_{i0},~~~
\pi_\theta=c_1 \dot\theta,
\label{momtt}
\end{equation}
the canonical Hamiltonian reads 
\begin{eqnarray}
{\cal H}^{(0)}&=&\frac{1}{2}\pi_i^2+\frac{1}{4}F_{ij}F^{ij}
+\frac{1}{2}m^2 (A^0)^2 +\frac{1}{2}m^2(A^i)^2
-A^0(\partial^i\pi_i+m^{2}A^{0})\nonumber\\
& &+\frac{1}{2c_1}\pi_\theta^2+\frac{1}{2}c_{1}(\partial_i\theta)^2
-c_2 A^{i}\partial_i\theta -c_3 f.
\end{eqnarray}
Here note that the equation for the canonical momentum $\pi_{0}$ in 
Eq. (\ref{momtt}) yields the constraint $\tilde{\Omega}_{1}$, which will 
be shown to be equivalent to the corresponding first-class constraint 
in Eq. (\ref{consttil}).

Following the canonical procedure for obtaining the equivalent symplectic 
first-order Lagrangian with the WZ term, we have
\begin{equation}
{\cal L}^{(0)}=\pi_{0}\dot{A}^{0}+\pi_i\dot{A}^i+\pi_\theta\dot\theta
-{\cal H}^{(0)},
\label{call0}
\end{equation}
where the initial set of symplectic variables $\xi^{(0)\alpha}$ and 
their conjugate momenta $a^{(0)}_\alpha$ are now given by
\begin{equation}
\xi^{(0)\alpha}=(\vec{A},\vec{\pi},\theta,\pi_\theta,A^0),~~~
a^{(0)}_\alpha=(\vec{\pi},\vec{0},\pi_\theta,0,-c_{2}\theta).
\label{a0wz}
\end{equation}
From Eq. (\ref{a0wz}) we read off the symplectic singular two-form matrix to be 
\begin{equation}
f^{(0)}_{\alpha\beta}(x,y)=
\left( 
\begin{array}{ccccc}
         O &-I  &\vec{0} &\vec{0}  &\vec{0}\\
         I &O   &\vec{0} &\vec{0}  &\vec{0}\\
         \vec{0}^{T} &\vec{0}^{T}   &0 &-1 &-c_{2}\\
         \vec{0}^{T} &\vec{0}^{T}   &1 &0  &0\\
         \vec{0}^{T} &\vec{0}^{T}   &c_{2} &0 &0
\end{array}
\right)\delta(x-y)
\label{f0wz0}
\end{equation}
having a non-trivial zero mode given by
\begin{equation}
\nu^{(0)T}_{\alpha,y}(1,x)=(\vec{0}, \vec{0}, 0, -c_{2}, 1)\delta (x-y).
\end{equation}
Applying this zero mode from the left to the equation of motion,
we are led to a constraint $\tilde{\Omega}_2$
\begin{equation}
\int {\rm d}^{3}y~ \nu^{(0)T}_{\alpha,y}(1,x) 
        \frac{\delta}{\delta\xi^{(0)\alpha}(y)} 
    \int {\rm d}^{3}z {\cal H}^{(0)}(z)
=-\tilde{\Omega}_2 (x) =0,
\label{nuomega0}
\end{equation}
where $\tilde{\Omega}_2$ is given by 
\begin{equation}
\tilde{\Omega}_{2}=\partial^{i}\pi_{i}+m^2 A^{0}
+\frac{c_{2}}{c_{1}}\pi_{\theta},
\label{constsym}
\end{equation}
which will be shown to be equal to the corresponding constraint in 
Eq. (\ref{consttil}) with the fixed values of $c_{1}$ and $c_{2}$.

Next, following the symplectic algorithm outlined before, we obtain the 
first-iterated Lagrangian by enlarging the canonical sector with the 
constraint $\tilde{\Omega}_2$ and its associated Lagrangian multiplier $\rho$ 
as follows 
\begin{equation}
{\cal L}^{(1)}=\pi_{0}\dot{A}^{0}+\pi_i\dot{A}^i+\pi_\theta\dot\theta
+\tilde{\Omega}_{2}\dot\rho-{\cal H}^{(1)}
\end{equation}
where ${\cal H}^{(1)}={\cal H}^{(0)}|_{\tilde{\Omega}_{2}=0}$ is the 
first-iterated Hamiltonian.  We have now for the first-level 
symplectic variables $\xi^{(1)\alpha}$ and their conjugate momenta 
$a^{(1)}_\alpha$ 
\begin{equation}
\xi^{(1)\alpha}=(\vec{A},\vec{\pi},\theta,\pi_\theta,A^0,\rho),~~~
a^{(1)}=(\vec{\pi},\vec{0},\pi_\theta,0,-c_{2}\theta,\tilde{\Omega}_2),
\end{equation}
and the first-iterated symplectic two-form matrix now reads 
\begin{equation}
f^{(1)}_{\alpha\beta}(x,y) =
\left( 
\begin{array}{cccccc}
  O &-I &\vec{0} &\vec{0}  &\vec{0} &\vec{0}\\
  I &O  &\vec{0} &\vec{0}  &\vec{0} 
  &-\vec{\nabla}_x \\
  \vec{0}^T & \vec{0}^T &0 &-1 &-c_{2} &0\\
  \vec{0}^T &\vec{0}^T  &1 &0  &0 &\frac{c_2}{c_1}\\
  \vec{0}^T &\vec{0}^T  &c_{2} &0  &0 &m^2 \\
  \vec{0}^T &\vec{\nabla}^{T}_{x} &0
         &-\frac{c_2}{c_1} &-m^{2} &0
\end{array}
\right)\delta (x-y).
\label{gif}
\end{equation}

In order to realize a gauge symmetry, this matrix must have at least one 
zero-mode which does not imply a new constraint.  To start out with, we 
introduce two zero-modes for the first-iterated symplectic two-form matrix 
\begin{eqnarray}
&&\nu^{(1)T}_{\alpha,y}(1,x)=(\vec{0}, \vec{0},0,-c_{2},1,0)\delta(x-y),
\nonumber\\
&&\nu^{(1)T}_{\alpha,y}(2,x)=(\vec{\nabla}_{x}, \vec{0}, 
-\frac{c_2}{c_1},0,0,1)\delta(x-y).
\label{zero1}
\end{eqnarray}
We require that these zero-modes should not generate any new constraint upon 
applying it from the left to the equation of motion
\begin{eqnarray}
\int {\rm d}^{3}y~\nu^{(1)T}_{\alpha,y}(1,x) \frac{\delta}
{\delta\xi^{(1)\alpha}(y)} \int {\rm d}^{3}z~{\cal H}^{(1)}(z) 
&=&(m^2-\frac{c_2^2}{c_1})A^{0},
\nonumber\\
\int {\rm d}^{3}y~\nu^{(1)T}_{\alpha,y}(2,x) \frac{\delta}
{\delta\xi^{(1)\alpha}(y)} \int {\rm d}^{3}z~{\cal H}^{(1)}(z) 
&=&(m^2-\frac{c_2^2}{c_1})\partial_{i}A^{i}+\frac{c_{2}c_{3}}{c_{1}}
\frac{df}{d\theta}.
\end{eqnarray}
Hence no new constraint is generated provided we choose for the 
free adjustable coefficients:
\begin{equation}
c_{1}=c_{2}=m^{2},~~~c_{3}=0.
\end{equation}
As a result, we arrived at the final result in the form of 
the St\"uckelberg Lagrangian (\ref{stuck}), which manifestly 
displays the gauge invariance under $\delta A^{\mu}=\partial^{\mu}\Lambda$ 
and $\delta\theta=-\Lambda$.  Note that with the above fixed $c_{i}$,  
$\tilde{\Omega}_{2}$ in Eq. (\ref{constsym}) is isomorphic to 
$\tilde{\Omega}_2$ given in Eq. (\ref{consttil}).

Now, in order to discuss gauge transformation, we consider the skew symmetric 
matrix (\ref{gif}) of the form,
\begin{equation}
f^{(1)}_{\alpha\beta}(x,y) =
\left( 
\begin{array}{cc}
        f_{\hat{\alpha}\hat{\beta}} &M_{\hat{\alpha}\sigma}\\
        -M^{T}_{\hat{\alpha}\sigma} &O  
\end{array}
\right)\delta(x-y),
\label{f1fab}
\end{equation}
where the submatrix $f_{\hat\alpha,\hat\beta}$ refers to
the $\xi_{\hat\alpha}=(\vec{A},\vec{\pi},\theta,\pi_\theta)$ sector, 
and $M_{\hat{\alpha}\sigma}$ is a $2\times 8$ matrix defined as  
$M_{\hat{\alpha}\sigma}\delta(x-y)=\frac{\partial\Omega_\sigma(y)}
{\partial \xi_{\hat{\alpha}}(x)}$.  Following ref.~\cite{wozneto}, 
the zero-modes $\nu^{(1)}_{\alpha,y}(\sigma,x)$ of $f^{(1)}(x,y)$ are 
of the general form
\begin{equation}
\nu^{(1)}_{\alpha,y}(1,x)=
\left(
\begin{array}{c}
-f^{-1}_{\hat{\alpha}\hat{\beta}}M_{\hat{\beta}1}\\ 
1\\
0\\
\end{array}
\right)\delta(x-y),~~~
\nu^{(1)}_{\alpha,y}(2,x)=
\left(
\begin{array}{c}
-f^{-1}_{\hat{\alpha}\hat{\beta}}M_{\hat{\beta}2}\\
0\\ 
1\\
\end{array}
\right)\delta(x-y).
\end{equation}
From Eq. (\ref{gif}) we have
\begin{equation}
f^{-1}_{\hat{\alpha}\hat{\beta}}=
\left( 
\begin{array}{cccc}
        O &I  &\vec{0} &\vec{0}     \\
       -I &O   &\vec{0} &\vec{0}     \\
        \vec{0}^{T}  &\vec{0}^{T}   &0  & 1  \\
        \vec{0}^{T}  &\vec{0}^{T}   &-1 &0 
\end{array}
\right),
\end{equation}
so that the zero-modes are (we have now $c_{2}/c_{1}=1$)
\begin{eqnarray}
&&\nu^{(1)T}_{\alpha,y}(1,x)=(\vec{0},\vec{0},0,-1,1,0)\delta(x-y),\nonumber\\
&&\nu^{(1)T}_{\alpha,y}(2,x)=(\vec{\nabla}_{x}, \vec{0}, 
               -1,0,0,1)\delta(x-y)
\label{zero2}
\end{eqnarray}
in agreement with Eq. (\ref{zero1}).

As has been shown in ref. \cite{wozneto}, the ``trivial'' zero modes
generate gauge transformation on $\xi^{(1)}_{\alpha}$ in the sense
\footnote{The Dirac algorithm as applied to the symplectic Lagrangian shows
that the gauge transformation on $\xi_{\hat{\alpha}}$ are generated by the 
first-class constraints $\Omega_\sigma(x)~(\sigma=1,2)$ with respect to
the symplectic Poisson bracket: 
$\delta\xi_{\hat{\alpha}}(x)=\{\xi_{\hat{\alpha}}(x),G\}_{symp}
=\int {\rm d}^{3}y~\frac{\partial \xi_{\hat{\alpha}}(x)}
{\partial \xi_{\hat{\beta}}(y)}f^{-1}_{\hat{\beta}\hat{\gamma}}
\frac{\delta G}{\delta \xi_{\hat{\gamma}}(y)}$ where 
$G=\Sigma_{\sigma}\int{\rm d}^3y~\epsilon_\sigma (y)\Omega_\sigma (y)$.  
Hence, we have 
$\delta\xi_{\hat{\alpha}}(x)=\Sigma_{\sigma} \int {\rm d}^3y 
f^{-1}_{\hat{\alpha}\hat{\gamma}}
\frac{\partial \Omega_\sigma(y)}{\partial \xi_{\hat{\gamma}}(x)}
\epsilon_{\sigma}(y)$ or the $\hat{\alpha}$-components of the zero-modes 
(\ref{zero2}) are seen to generate the gauge transformation on 
$\xi_{\hat{\alpha}}$.}
\begin{equation}
\delta\xi_{\alpha}(x)=\Sigma_\sigma 
        \int {\rm d}^3y~\nu^{(1)T}_{\alpha,y}(\sigma,x)\epsilon_\sigma(y).
\label{trfmrule}
\end{equation}
we thus conclude from Eq. (\ref{zero2}),
\begin{eqnarray}
\delta A^0 &=& \epsilon_1,~~~
\delta A^i = \partial^i \epsilon_2,~~~
\delta \theta = -\epsilon_2,\nonumber\\
\delta\pi_i &=&0,~~~ 
\delta\pi_\theta=-m^{2}\epsilon_{1},~~~ 
\delta\rho=\epsilon_2.
\end{eqnarray}
As one readily checks, these only represent a symmetry transformation
of the symplectic first level Lagrangian ${\cal L}^{(1)}$
if $\epsilon_1=\dot{\epsilon}_2$. In that case it is also a symmetry
of the St\"uckelberg Lagrangian (\ref{stuck}). As was shown in ref.~\cite{brr},
this condition also follows from the requirement that the gauge transformation
commutes with the time-derivative operation in Hamilton's equations of
motion.

\begin{center}
{\bf BRST invariant symplectic embedding}
\end{center}

Following the Batalin-Fradkin-Vilkovisky 
formalism~\cite{fradkin75,henneaux85} in 
the extended phase space, we introduce a minimal set of the ghost and 
antighost fields together with auxiliary fields as follows
\beq
({\cal C},\bar{\cal P}),~~({\cal P},\bar{\cal C}),~~
(N,B).
\eeq
Similar to the previous WZ action case, we now construct the 
BRST invariant gauge-fixed Lagrangian in the symplectic scheme by including 
the above auxiliary fields with ghost terms in the Lagrangian,
\begin{eqnarray}
{\cal L}&=&{\cal L}_{0}+{\cal L}_{WZ}+{\cal L}_{GF},\nonumber\\
{\cal L}_{0}+{\cal L}_{WZ}&=&-\frac{1}{4}F_{\mu\nu}F^{\mu\nu}
+\frac{1}{2}m^2 (A_\mu + \partial_\mu\theta)(A^\mu + \partial^{\mu}\theta),
\end{eqnarray}
where we take an Ansatz for the gauge-fixing (GF) Lagrangian respecting 
Lorentz symmetry
\begin{equation}
{\cal L}_{GF}=d_1 A^{\mu}\partial_\mu B+d_{2}\partial_{\mu}\bar{\cal C}
\partial^{\mu}{\cal C}+d_{3}B^{2}+ d_{4}g.
\label{gflag}
\end{equation}
with the $\theta$-dependent function $g$.  Here note that we have used the 
canonical momentum field $B$ instead of the multiplier field $N$ in this 
Ansatz to construct the desired well-known ghost Lagrangian in the Proca 
model.  Through the Legendre transformation we can obtain the canonical 
Hamiltonian of the form
\begin{eqnarray}
{\cal H}^{(0)}&=&\frac{1}{2}\pi_i^2+\frac{1}{4}F_{ij}F^{ij}
+\frac{1}{2}m^2 (A^0)^2 +\frac{1}{2}m^2(A^i)^2
-A^0(\partial^i\pi_i+m^{2}A^{0})\nonumber\\
& &+\frac{1}{2m^{2}}\pi_\theta^2+\frac{1}{2}m^{2}(\partial_i\theta)^2
-m^{2}A^{i}\partial_i\theta -(d_{1}-1)A^{\mu}\partial_{\mu}B
-\frac{1}{d_{2}}\bar{\cal P}{\cal P}\nonumber\\
& &+d_{2}\partial_{i}\bar{\cal C}\partial_{i}{\cal C}
-d_{3}B^{2}-d_{4}g 
\label{h0brst}
\end{eqnarray}
where the canonical momenta conjugate to $A^0$, $A^i$, $\theta$, ${\cal C}$ 
and $\bar{\cal C}$ are given as
\begin{equation}
\pi_0=-m^{2}\theta,~~~
\pi_i=F_{i0},~~~
\pi_{\theta}=m^{2}\dot{\theta},~~~
\bar{\cal P}=d_{2}\dot{\bar{\cal C}},~~~
{\cal P}=-d_{2}\dot{\cal C},
\label{momtt2}
\end{equation} 
and satisfy the super-Poisson algebra
\footnote{Here the super-Poisson bracket is defined as 
$\{A,B\}=\frac{\delta A}{\delta q}|_{r}\frac{\delta B}{\delta p}|_{l}
-(-1)^{\eta_{A}\eta_{B}}\frac{\delta B}{\delta q}|_{r}\frac{\delta A} {%
\delta p}|_{l} $ where $\eta_{A}$ denotes the ghost number in $A$
and the subscript $r$ and $l$ imply right and left derivatives, 
respectively.}
$$
\{{\cal C}(x),\bar{{\cal P}}(y)\}=\{{\cal P}(x), 
\bar{{\cal C}}(y)\}=\{N(x),B(y)\}=\delta(x-y).
$$
Here note that since in the GF Lagrangian (\ref{gflag}) we have  already 
introduced the momenta field $B$, we do not have any specific relation 
between $B$ and $N$ in Eq. (\ref{momtt2}), and thus we still have the covariant 
term proportional to $A^{\mu}\partial_{\mu}B$ in Eq. (\ref{h0brst}), where 
we have also used the identification\footnote{
Note that, differently from the identification $N=-A^{0}$ in the 
literature~\cite{kkky}, here we have used a highly nontrivial relation.}
\begin{equation}
\dot{N}=-\partial_{\mu}A^{\mu},
\label{ndot}
\end{equation}
which plays a crucial role in construction of the BRST symmetry and 
is also related to the integrability contidion in Eqs. 
(\ref{const3}) and (\ref{h3p}) in the next section.

Following the canonical procedure for obtaining the symplectic 
first-order Lagrangian, we have
\begin{equation}
{\cal L}^{(0)}=\pi_{0}\dot{A}^{0}+\pi_i\dot{A}^i+\pi_\theta\dot\theta
+B\dot{N}+\bar{\cal P}\dot{\cal C}+\bar{\cal C}\dot{\cal P}
-{\cal H}^{(0)},
\label{call0brst}
\end{equation}
where the initial set of symplectic variables $\xi^{(0)\alpha}$ and 
their conjugate momenta $a^{(0)}_\alpha$ are now given by
\begin{eqnarray}
\xi^{(0)\alpha}&=&(\vec{A},\vec{\pi},\theta,\pi_\theta,A^0,N,B,{\cal C}, 
\bar{\cal P},\bar{\cal C},{\cal P}),\nonumber\\
a^{(0)}_\alpha&=&(\vec{\pi},\vec{0},\pi_\theta,0,-m^{2}\theta,B,0,\bar{\cal P},0,{\cal P},0).
\label{a0brst}
\end{eqnarray}
From Eq. (\ref{a0brst}) we read off the symplectic singular two-form matrix 
to be 
\begin{equation}
f^{(0)}_{\alpha\beta}(x,y)=
\left( 
\begin{array}{cc}
         f_{\hat{\alpha}\hat{\beta}}  &O\\
         O   &f^{GF}_{\mu\nu}
\end{array}
\right)\delta(x-y),
\end{equation}
where $f_{\hat{\alpha}\hat{\beta}}$ is a $9\times 9$ submatrix, 
which can be read off from Eq. (\ref{f0wz0}), and $f^{GF}_{\mu\nu}$ is 
a $6\times 6$ submatrix defined as
\begin{equation}
f^{GF}_{\mu\nu}=
\left( 
\begin{array}{ccc}
         -\epsilon &O          &O\\
         O         &-\epsilon  &O\\
         O         &O          &-\epsilon

\end{array}
\right),
\end{equation}
with $\epsilon$ being the Levi-Civita tensor with $\epsilon_{12}=1$ and 
$O$ being the $2\times 2$ null matrix.  As in Eq. (\ref{nuomega0}), 
using a non-trivial zero mode 
\begin{equation}
\nu^{(0)T}_{\alpha,y}(1,x)=(\vec{0}, \vec{0}, 0, -m^{2}, 1, 
0, 0, 0, 0, 0, 0)\delta (x-y).
\end{equation}
we obtain a constraint $\tilde{\Omega}_2$
\begin{equation}
\tilde{\Omega}_{2}=\partial^{i}\pi_{i}+m^2 A^{0}+\pi_{\theta}+(d_{1}-1)\dot{B},
\label{constsymbrst}
\end{equation}
which will be shown to be equal to the corresponding constraint in 
Eq. (\ref{consttil}) with the fixed values of $d_{1}$.
 
Next, as in the gauge invariant symplectic embedding case, we obtain the 
first-iterated Lagrangian by enlarging the canonical sector with the 
constraint $\tilde{\Omega}_2$ and its associated Lagrangian multiplier $\rho$ 
\begin{equation}
{\cal L}^{(1)}=\pi_{0}\dot{A}^{0}+\pi_i\dot{A}^i+\pi_\theta\dot\theta
+B\dot{N}+\bar{\cal P}\dot{\cal C}+\bar{\cal C}\dot{\cal P}
+\tilde{\Omega}_{2}\dot\rho-{\cal H}^{(1)}
\end{equation}
where ${\cal H}^{(1)}={\cal H}^{(0)}|_{\tilde{\Omega}_{2}=0}$ is the 
first-iterated Hamiltonian.  We now obtain the first-level 
symplectic variables $\xi^{(1)\alpha}$ and their conjugate momenta 
$a^{(1)}_\alpha$
\begin{eqnarray} 
\xi^{(0)\alpha}&=&(\vec{A},\vec{\pi},\theta,\pi_\theta,A^0,N,B,{\cal C}, 
\bar{\cal P},\bar{\cal C},{\cal P},\rho),\nonumber\\
a^{(0)}_\alpha&=&(\vec{\pi},\vec{0},\pi_\theta,0,-m^{2}\theta,B,0,
\bar{\cal P},0,{\cal P},0,\tilde{\Omega}_{2}).
\label{a1brst}
\end{eqnarray}
and the first-iterated symplectic two-form matrix  
\begin{equation}
f^{(1)}_{\alpha\beta}(x,y)=
\left( 
\begin{array}{ccc}
         f_{\hat{\alpha}\hat{\beta}}  &O &\vec{m}\\
         O   &f^{GF}_{\mu\nu} &\vec{m}_{GF}\\
        -\vec{m}^{T} &-\vec{m}_{GF}^{T} &0
\end{array}
\right)\delta(x-y),
\end{equation}
with $\vec{m}^{T}=(\vec{0},-\vec{\nabla}_{x},0,1,m^{2})$ and 
$\vec{m}_{GF}^{T}=(0,-(d_{2}-1)\partial_{t},0,0,0,0)$.

In order to realize the BRST symmetry, we introduce two zero-modes
\begin{eqnarray}
\nu^{(1)T}_{\alpha,y}(1,x)&=&(\vec{0}, 
\vec{0},0,-m^{2},1,0,0,0,0,0,0,0)\delta(x-y),
\nonumber\\
\nu^{(1)T}_{\alpha,y}(2,x)&=&(\vec\nabla_{x}, \vec{0}, 
-1,0,0,-(d_{2}-1)\partial_{t},0,0,0,0,0,1)\delta(x-y).
\label{zero1brst}
\end{eqnarray}
We require that these zero-modes should not generate any new constraint upon 
applying it from the left to the equation of motion
\begin{equation}
\int {\rm d}^{3}y~\nu^{(1)T}_{\alpha,y}(2,x) \frac{\delta}
{\delta\xi^{(1)\alpha}(y)} \int {\rm d}^{3}z~{\cal H}^{(1)}(z) 
=(d_{1}-1)\partial_{i}\partial^{i}B+d_{4}\frac{\partial g}{\partial\theta},
\end{equation}
and the equation corresponding to the zero-mode 
$\nu^{(1)T}_{\alpha,y}(1,x)$ yields trivial identity.  In order to 
guarantee no new constraint, we choose for the free adjustable coefficients:
\begin{equation}
d_{1}=1,~~~d_{3}=-\frac{1}{2}\alpha,~~~d_{4}=0,
\end{equation}
where we have used the conventional form for $d_{3}$ to be consistent with 
the BRST gauge fixing term.  As a result, we have arrived at the Lorentz 
invariant Lagrangian of the form
\begin{equation}
{\cal L}=- \frac{1}{4}  F_{\mu\nu} F^{\mu\nu}
               + \frac{1}{2} m^2 (A_\mu + \partial_\mu \theta)^{2}
+A^{\mu}\partial_\mu B+d_{2}\partial_{\mu}\bar{\cal C}
\partial^{\mu}{\cal C}-\frac{1}{2}\alpha B^{2}. 
\label{brstlag}
\end{equation}
Exploiting the transformation rules (\ref{trfmrule}), with the generalized 
zero-modes $\nu^{(1)T}_{\alpha,y}(\sigma,x)$ in Eq. (\ref{zero1brst}), the 
replacement of $\epsilon_{2}=-\lambda{\cal C}$ and an additional 
BRST transformation rule for the $\bar{\cal C}$, we obtain the BRST 
transformation rules having nilpotent property $\delta_{B}^{2}=0$ as follows,
\begin{equation}
\delta_{B}A^{\mu}=-\lambda\partial^{\mu}{\cal C},~~~
\delta_{B}\theta =\lambda{\cal C},~~~
\delta_{B}\bar{\cal C}=-\lambda B,~~~
\delta_{B}{\cal C}=\delta_{B}B=0,
\label{brsttrfm}
\end{equation}
under which we obtain
\begin{equation}
\delta_{B}{\cal L}=-(d_{2}+1)\lambda\partial_{\mu}{\cal C}\partial^{\mu}B.
\label{brstvariation}
\end{equation}
With the fixed value of $d_{2}$:
\begin{equation}
d_{2}=-1
\label{d2}
\end{equation}
we have finally obtained the desired BRST invariant Lagrangian
\begin{equation}
{\cal L}=- \frac{1}{4}  F_{\mu\nu} F^{\mu\nu}
+\frac{1}{2}m^{2}(A_{\mu}+\partial_{\mu}\theta)
(A^{\mu}+\partial^{\mu}\theta)
+A^{\mu}\partial_\mu B-\partial_{\mu}\bar{\cal C}
\partial^{\mu}{\cal C}-\frac{1}{2}\alpha B^{2}. 
\label{brstlag2}
\end{equation}
Note that in this symplectic formalism, one can avoid algebra of complicated 
structure having fermionic gauge fixing function and minimal Hamiltonian 
needed in the standard BRST formalism~\cite{kkky}. 

\section{Hamilton-Jacobi analysis}
\setcounter{equation}{0}
\renewcommand{\theequation}{\arabic{section}.\arabic{equation}}

In this section we apply the HJ method~\cite{guler89,pimentel98,hong01ph} 
to the Proca model where the generalized HJ PDEs are given as 
\begin{equation}
\label{hj}
{\cal H}^{\prime}_0 =p_{0} + {\cal H}_{0} =0,~~~
{\cal H}^{\prime}_1 =\pi_0 +{\cal H}_{1}=0,
\end{equation}
where ${\cal H}_{0}$ is a canonical Hamiltonian density (2.2) and 
${\cal H}_{1}$ is actually zero for the Proca case.
Note that ${\cal H}_{1}^{\pr}$ is the primary constraint in the Dirac 
terminology~\cite{dirac64}.  From Eq. (\ref{hj}) one obtains after some 
calculation, 
\begin{equation}
\label{hjem}
dq_{\un{\mu}} =\frac{\pa {\cal H}^{\prime}_{\un{\alpha}}}
{\pa p_{\un{\mu}}}dt_{\un{\alpha}},~~~
dp_{\un{\mu}}=-\frac{\pa {\cal H}^{\prime}_{\un{\alpha}}}
{\pa q_{\un{\mu}}}dt_{\un{\alpha}},
\end{equation}
where $q_{\un{\mu}}=(t,A^{0},A^{i})$, $p_{\un{\mu}}=(p_{0},\pi_{0},\pi_{i})$ 
and $t_{\un{\alpha}}=(t,A^{0})$.

Exploiting the Hamilton equations (\ref{hjem}), we obtain the set of 
equations of motion
\begin{eqnarray}
&& dA_0=dA_0,~~~dA^i=(\pi_i+\partial^iA^0)dt, \nonumber \\
&& d\pi_0=(\partial^i\pi_i+m^2A^0)dt, 
   ~~~d\pi_i=(\partial_j F^{ij}-m^2A^{i})dt.
\label{da0dpi0}
\end{eqnarray}
Note that, since the equation for $A^0$ is trivial, one cannot obtain any
information about the variable $A^0$ at this level, and the set of equation 
is not integrable.  

In order to remedy these unfavorable problems, we have to investigate the 
integrability conditions,\footnote{Note that even though
${\cal H}^{\prime}_{\bar{\alpha}}$ carry the extended index $\bar{\alpha}$ 
($\bar{\alpha}=0,1,2,3)$ with the additional constraints, the coordinates 
$t_{\un{\alpha}}$ carry only the index $\un{\alpha}$ since one cannot 
generate coordinates themselves.} 
\begin{equation}
\dot{\cal H}^{\prime}_{\bar{\alpha}}
=\{{\cal H}^{\prime}_{\bar{\alpha}}, {\cal H}^{\prime}_{0}\}
+\{{\cal H}^{\prime}_{\bar{\alpha}},{\cal H}^{\prime}_{\beta}\}
\dot{q}_{\beta}=0,
\label{integ2}
\end{equation}
where, unlike in the usual case, the Poisson bracket is defined as
\begin{equation}
\label{partial}
\{A,B\}
=\frac{\pa A}{\pa q_{\un{\mu}}}\frac{\pa B}
{\pa p_{\un{\mu}}}
-\frac{\pa B}{\pa q_{\un{\mu}}}\frac{\pa A}
{\pa p_{\un{\mu}}},
\end{equation}
with the extended index $\un{\mu}$ corresponding to 
$q_{\un{\mu}}=(t,A^{0},A^{i})$.  Eq. (\ref{integ2}) then imply that
\begin{equation}
\dot{\cal H}'_0 = -{\cal H}'_2,~~~\dot{\cal H}'_1 = {\cal H}'_2, 
\label{const2}
\end{equation}
where ${\cal H}_{2}^{\prime}$ is a secondary
constraint in Dirac terminology given as
\begin{equation}
{\cal H}'_2=\partial^i\pi_i+m^2A^0=0.
\label{h2}
\end{equation}
This ${\cal H}'_2$ then yields an additional integrability condition as
\beq
\dot{\cal H}^{\pr}_{2}=m^2{\cal H}^{\pr}_{3},
\label{const3}
\eeq
where
\beq
{\cal H}_{3}^{\pr}=\partial_i A^i+\dot{A}_0=0,
\label{h3p}
\eeq
which provides the missing information for the Hamilton
equations for $A_0$. As a result, one can easily get the desired
equations of motion (\ref{da}).  Moreover, in Eqs. (\ref{const3}) 
and (\ref{h3p}), time evolution of ${\cal H}_{2}^{\prime}$ can 
be rewritten in the nontrivial covariant form: 
$\dot{\cal H}^{\pr}_{2}=m^{2}\partial_{\mu}A^{\mu}$ and such somehow unusual 
structure has been already seen in Eq. (\ref{ndot}) in the symplectic 
embedding.  Note that ${\cal H}_{3}^{\pr}$ yields the value of $\dot{A}^{0}$ 
which is exactly the same as the above fixed value of $u$, and also the Poisson brackets in the HJ scheme are 
the same as those in the DQM since $\Omega_{i}$ do not depend on time 
explicitly.  Moreover, if the generalized HJ PDEs in Eqs. (\ref{hj}) and 
(\ref{h2}) are rewritten in terms of $\Omega_{1}(={\cal H}_{1}^{\prime})$ and 
$\Omega_{2}(={\cal H}_{2}^{\prime})$ and $u(=\dot{A}_{0})$, one can easily 
reproduce the integrability conditions in Eqs. (\ref{const2}) and 
(\ref{const3}), thus showing that the integrability conditions in HJ scheme 
are equivalent to the consistency conditions in DQM. 

Now we consider the closeness of the Lie algebra involved in the HJ scheme 
by introducing operators  $X_{\un{\alpha}}$
($\un{\alpha}=0,1$) corresponding to ${\cal H}_{\un{\alpha}}^{\prime}$, 
formally defined by
\begin{equation}
\label{partialx}
X_{\un{\alpha}} f=\frac{\pa f}{\pa t_{\un{\alpha}}}
+\frac{\pa f}{\pa q_{i}}\frac{\pa {\cal H}_{\un{\alpha}}^{\prime}}{\pa p_{i}}
-\frac{\pa f}{\pa p_{i}}\frac{\pa {\cal H}_{\un{\alpha}}^{\prime}}{\pa q_{i}}.
\end{equation}
Using Eq. (\ref{partialx}) we then obtain 
\bea
X_{0}f&=& \frac{\pa f}{\pa t}+(\pi_{i}+\pa^{i}A^{0})\frac{\delta f}
{\delta A^{i}}+(\pa_{j}F^{ij}-m^{2}A^{i})\frac{\delta f}{\delta \pi_{i}},
\nonumber\\
X_{1}f&=&\frac{\delta f}{\delta A^{0}},
\label{partialx1x2}
\eea
to yield the commutator relation among the operators $X_{\un{\alpha}}$
\beq
[X_{0},X_{1}]f=-\left(\pa_{i}\frac{\delta f}{\delta A^{i}}\right)\delta(x-y).
\eeq
Since the above commutator relation is not closed, we need to extend 
the set $\{X_{\un{\alpha}}\}$ to a set of operators $\{X_{\bar{\alpha}}\}$ 
($\bar{\alpha}=0,1,2,3)$ by introducing new operators.  In fact, after some 
algebra, we can construct two new operators $X_{2}$ and $X_{3}$:
\bea
X_{2}f&=&\pa_{i}\frac{\delta f}{\delta A^{i}},\nonumber\\
X_{3}f&=&\pa_{i}\frac{\delta f}{\delta \pi_{i}},
\label{x2x3ops}
\eea
to yield a closed Lie algebra
\bea
& &[X_{0},X_{1}]f=-X_{2}f\delta (x-y),~~~
[X_{0},X_{2}]f=-m^{2}X_{3}f\delta (x-y),
\nonumber\\
& &[X_{0},X_{3}]f=[X_{1},X_{2}]f=[X_{1}, X_{3}]f=[X_{2}, X_{3}]f=0,
\label{lies}
\eea
which automatically guarantee the integrability conditions discussed above.

Finally, we discuss the integrability conditions in terms of action.  
In fact, Eq. (\ref{hjem}) yields
\begin{equation}
\label{hjem1}
dS=dt_{\un{\alpha}}\int{\rm d}^{3}x~\left(-{\cal H}_{\un{\alpha}}
+\pi_{i}\frac{\delta {\cal H}^{\prime}_{\un{\alpha}}}
{\delta \pi_{i}}\right),
\end{equation}
from which we have obtained the action of the form
\begin{equation}
S=\int{\rm d}^{3}x~ \left(-{\cal H}_{0}dt+\pi_{0}dA^{0}
+\pi_{i}dA^{i}\right).
\label{action0ii}
\end{equation}
Since we can now have full equations of motion for $A^{i}$ and $A^{0}$ from  
Eqs. (\ref{da0dpi0}) and (\ref{h2}), respectively, $dA^{\mu}$ can 
be integrable to yield the desired expression 
$dA^{\mu}=\dot{A}^{\mu}dt$.  We can thus arrive at the desired 
standard action
\beq
S=\int{\rm d}^{4}x~ {\cal L}_{0}
\label{action0fin}
\eeq
where ${\cal L}_{0}$ is exactly the same as the first-order Lagrangian 
(\ref{lagfirst}) in the symplectic formalism.  Note that it was through 
the introduction of the secondary constraint obtained by the integrability 
condition (\ref{integ2}) that one could construct the action 
(\ref{action0fin}) even in the second-class system. 

\section{Conclusion}

It has been the primary objective of this paper to demonstrate in form of
a simple model how the improved Dirac quantization method (DQM) program of
embedding second-class Hamiltonian systems into first-class ones in the
context of Dirac's quantization procedure can be realized in the framework of
the symplectic approach to constrained systems.  Rather than proceeding
iteratively as in the improved DQM approach, we have greatly simplified the
calculation by making use of manifest Lorentz invariance in our {\it Ansatz}
for the Wess-Zumino (WZ) term. Just as in the case of the improved DQM 
procedure, this {\it Ansatz} clearly shows, that the embedding procedure 
requires the introduction of an even number of additional fields, which, 
following the Faddeev-Jackiw prescription~\cite{jackiw85} can be chosen to be
canonically conjugate pairs.  Indeed, the number of second-class constraints
is always even, and we know from the improved DQM embedding procedure that
phase space must be augmented by one degree of freedom for each secondary
constraint.  This fact has not been recognized in a recent paper on this
subject~\cite{neto0109} where in our notation $\pi_{\theta}$ has been taken
to be a function of $A^{i}$, $\pi_{i}$ and  $\theta$.  Correspondingly they
did not obtain the desired Wess-Zumino Lagrangian.  On the other hand, our
procedure treating $\pi_{\theta}$ as an independent field, led in a natural
way to the St\"uckelberg Lagrangian, in agreement with earlier
calculations using the improved DQM~\cite{kimjkps2}.  Similar to the WZ term 
case, we introduced the additional ghost and antighost fields together with 
the auxiliary fields in the symplectic scheme to construct the BRST invariant 
gauge-fixed Lagrangian and its nilpotent BRST transformation rules.  In this 
approach we could avoid complicated calculations of the minimal Hamiltonian 
associated with the fermionic gauge fixing function.

Finally, we have also demonstrated explicitly for the model in question that
the ``integrability conditions" of the Hamilton-Jacobi (HJ) scheme are just the
``consistency conditions" of the standard Dirac quantization procedure.  Next, 
we have constructed operators in terms of the generalized Poisson brackets, 
to obtain the closed Lie algebra associated with the commutator relations 
among these operators, and to guarantee the integrability conditions using 
the closed Lie algebra.  Moreover, using the integrability conditions in the 
HJ formalism, we have reconstructed the desired standard action, equivalent to 
the first-order Lagrangian in the symplectic scheme.  

\acknowledgments

One of us (KDR) would like to thank the Sogang High Energy
Physics Group for their warm hospitality and  acknowledges 
financial support from DFG-KOSEF Exchange Program. 
Two of us (STH and YJP) acknowledge financial support from 
the Korea Research Foundation, Grant No. KRF-2001-DP0083.

\end{document}